\documentstyle[aps,epsfig]{revtex}

\newcommand{\mt}[1]{\tiny #1}

\newcommand{\barq}{\langle\bar{q}q\rangle}
\newcommand{\abarq}{\langle\bar{q}i\gamma_5q\rangle}
\newcommand{\pe}{p_{\tiny{E}}}

\newcommand{\dtmu}{{\cal D}_{T,\mu}(p)}
\newcommand{\stmu}{\Sigma_{T,\mu}(p)}

\newcommand{\ptmu}{\Sigma_{5;T,\mu}(p)}

\newcommand{\gev}{\mbox{\small GeV}}

\begin{document}
\tightenlines
\draft

\title{Chiral meson masses at the critical point\\
from QCD in the improved ladder approximation}
\author{O. Kiriyama\thanks{E-mail: kiriyama@rcnp.osaka-u.ac.jp}}
\address{Research Center for Nuclear Physics, Osaka University,
Ibaraki 567--0047, Japan}
\maketitle

\begin{abstract}
Chiral meson masses at the critical point is investigated 
using QCD in the improved ladder approximation. 
We calculate the effective potential at finite temperature $T$ 
and quark chemical potential $\mu$ and find the critical point at 
$T \simeq 95$ MeV, $\mu \simeq 290$ MeV. 
The chiral meson masses are determined from a 
second derivative of the effective potential at its minimum. 
We find that the sigma goes to massless at the critical point 
while pion remains massive. 
Our results are consistent with that of the linear sigma model 
in contrast to the Nambu--Jona-Lasinio model.
\end{abstract}


Recently there has been great interest in studying the phase structure of 
quantum chromodynamics (QCD). We expect that at sufficiently high temperature 
and/or density the QCD vacuum undergoes a phase transition 
into a chirally symmetric/deconfinement phase\cite{MCLERRAN}, 
a color superconducting phase
\cite{CSC}. They may be realized in high-energy heavy-ion collisions at 
the BNL Relativistic Heavy Ion Collider (RHIC) 
and CERN Large Hadron Collider (LHC). 
These phase transitions are also important in the physics 
of neutron (or quark) stars and the early universe. 

In massless two--flavor QCD, as confirmed by using 
the lattice simulation\cite{EJIRI} 
and several effective theories
\cite{NJL,SCAVE,QCD,TANIGUCHI,HARADA,KIRIYAMA,IKEDA}, 
the chiral phase transition at high temperature is probably second order. 
On the other hand, at high density a first order one is expected
\cite{NJL,SCAVE,QCD,TANIGUCHI,HARADA,KIRIYAMA,IKEDA}. 
These observations indicate the existence of a tricritical point 
in the $\mu$-$T$ phase diagram for zero current quark masses 
and a critical point, where the first order transition ends, 
for nonzero current quark masses. 
The existence and the location of the critical point has been 
studied also by using the recently proposed 
lattice QCD method \cite{EP}. 
It is expected that the sigma field becomes massless at the critical point 
while the pion field remains massive and the critical point 
is in the same universality class as the three dimensional 
Ising model\cite{ISING}. Recently, it has been proposed 
that this point may lead to 
characteristic signatures which enable us to explore 
the phase structure of QCD in heavy-ion collisions\cite{CSD}. 
The behavior of the chiral meson masses around the critical point 
is studied within the framework of 
the linear sigma model and Nambu--Jona-Lasinio (NJL) model\cite{SCAVE}. 
In the linear sigma model, the sigma mass goes to zero at the critical 
point. However, in the NJL model with the random phase approximation 
the sigma remains massive at the critical point. 
The chiral meson masses at finite temperature and density 
has been studied in another QCD motivated model\cite{BARDUCCI}, 
however, their behavior around the critical point has not been known yet. 
In this paper we investigate where the critical point locates and 
how the chiral meson masses behave around the critical point 
using the so-called {\it QCD in the improved ladder approximation}
\cite{TANIGUCHI,HARADA,KIRIYAMA,IKEDA}. 

In the following we fix the scale parameter of our model by the condition 
$\Lambda_{\mt{QCD}}=1$ except for numerical calculations. 
Its value is determined by the condition $f_\pi=93$ MeV at $T=\mu=0$.

At zero temperature and zero chemical potential, 
the Cornwall-Jackiw-Tomboulis effective potential\cite{CJT} 
for QCD in the improved ladder approximation 
is expressed as a functional of $\Sigma(\pe)$ and $\Sigma_5(\pe)$
\cite{KIRIYAMA}, the scalar and the pseudo-scalar part of 
the dynamical mass function of the quark respectively
\begin{eqnarray}
V&=&-2\int\frac{d^4\pe}{(2\pi)^4}~
\ln\frac{\Sigma^2(\pe)+\Sigma_5^2(\pe)+\pe^2}{\pe^2}\nonumber\\
&&-\frac{2}{3C_2}\int d\pe^2~\frac{1}{\Delta(\pe)}
\left[\left(\frac{d}{d\pe^2}\Sigma(\pe)\right)^2
+\left(\frac{d}{d\pe^2}\Sigma_5(\pe)\right)^2\right],
\label{eqn:vcjt2}
\end{eqnarray}
where the function
\begin{eqnarray}
\Delta(\pe)=\frac{d}{d\pe^2}\frac{\bar{g}^2(\pe)}{\pe^2}
\end{eqnarray}
is introduced, $\pe$ denotes the Euclidean momentum, 
$\bar{g}^2(\pe)$ is the QCD running coupling 
of one-loop order, $C_2$ is the quadratic Casimir operator for 
color $SU(N_c)$ group, and an overall factor 
(the number of light flavors times the number of colors) is omitted. 
Note that in the derivation of Eq. (\ref{eqn:vcjt2}), 
the Higashijima--Miransky approximation\cite{HIG,MIR} has been used 
and an infrared finite running coupling and 
quark mass functions like Eqs. (\ref{eqn:ec}), 
(\ref{eqn:mass}) and (\ref{eqn:mass5}) are assumed. 
In this paper, we use the following effective running coupling\cite{HIG}
\begin{eqnarray}
\bar{g}^2(\pe)=\frac{2\pi^2a}{\ln (\pe^2+p_R^2)},\label{eqn:ec}
\end{eqnarray}
where $p_R$ is an infrared regularization parameter to keep 
the QCD running coupling from blowing up at $\pe=1(\Lambda_{\mt{QCD}})$, 
and $a=8/9$ (we use three--flavor, three--color effective running coupling). 
Corresponding to the above running coupling, 
the Schwinger--Dyson equations, which are the extremum conditions 
of Eq. (\ref{eqn:vcjt2}) with respect to $\Sigma(\pe)$ or $\Sigma_5(\pe)$, 
suggest the following trial mass functions
\begin{eqnarray}
\Sigma(\pe)&=&m_R[\ln (\pe^2+p_R^2)]^{-a/2}
+\frac{\sigma}{\pe^2+p_R^2}[\ln (\pe^2+p_R^2)]^{a/2-1},\label{eqn:mass}\\
\Sigma_5(\pe)&=&\frac{\sigma_5}{\pe^2+p_R^2}
[\ln (\pe^2+p_R^2)]^{a/2-1}\label{eqn:mass5},
\end{eqnarray}
where $\sigma$ ($\sigma_5$) is related to the 
renormalization group invariant (RGI) quark condensate 
$\barq$ ($\abarq$) as $\sigma=2\pi^2a\barq/3$ 
($\sigma_5=2\pi^2a\abarq/3$) and $m_R$ is the RGI current quark mass. 
The explicit chiral symmetry breaking term is introduced 
in $\Sigma(\pe)$ alone as it should be.

We now turn to discussions about the effective potential 
at finite temperature and density. 
In order to calculate the effective potential at finite 
temperature and density we apply the imaginary time formalism\cite{JIK}
\begin{eqnarray}
\int\frac{dp_4}{2\pi}f(p_4) \to 
T\sum_{n=-\infty}^{\infty}f(\omega_n+i\mu),\label{eqn:itf}
\end{eqnarray}
where $\omega_n=(2n+1)\pi T$ $(n=0,\pm 1,\pm 2,\cdots)$ 
is the fermion Matsubara frequency 
and $\mu$ represents the quark chemical potential. 
In addition, we need to define the running coupling 
and the trial mass functions at finite temperature and density. 
For simplicity, we adopt the following functional form 
for $\dtmu$, $\stmu$, and $\ptmu$ 
by replacing $p_4$ in ${\cal D}(\pe)$, $\Sigma(\pe)$, 
and $\Sigma_5(\pe)$with $\omega_n$:
\begin{eqnarray}
\dtmu&=&\frac{2\pi^2a}{\ln(\omega_n^2+|\vec{p}|^2+p_R^2)}~
\frac{1}{\omega_n^2+|\vec{p}|^2},\label{eqn:dp}\\
\stmu&=&m_R\left[\ln(\omega_n^2+|\vec{p}|^2+p_R^2)\right]^{-a/2}\nonumber\\
&&+\frac{\sigma}{\omega_n^2+|\vec{p}|^2+p_R^2}
\left[\ln(\omega_n^2+|\vec{p}|^2+p_R^2)\right]^{a/2-1}\label{eqn:sigma},\\
\ptmu&=&\frac{\sigma_5}{\omega_n^2+|\vec{p}|^2+p_R^2}
\left[\ln(\omega_n^2+|\vec{p}|^2+p_R^2)\right]^{a/2-1}\label{eqn:pi}.
\end{eqnarray}
A few comments are in order.

In Eq. (\ref{eqn:dp}) we do not introduce the $\mu$ dependence 
in $\dtmu$. 
The gluon momentum squared is the most natural argument of the running 
coupling at zero temperature and density, in the light of the chiral 
Ward-Takahashi identity\cite{JAIN,KUGO}. 
Then it is reasonable to assume that $\dtmu$ 
does not depend on the quark chemical potential. 
In addition, the screening mass is not included 
in Eq. (\ref{eqn:dp}). 
As concerns the mass functions, we use the same function 
as Eqs. (\ref{eqn:mass}) and (\ref{eqn:mass5}) 
except that we replace $p_4$ with $\omega_n$. 
The quark wave function does not suffer the 
renormalization in the Landau gauge for $T=\mu=0$, while, the same does not 
hold for finite temperature and/or density. 
Furthermore, we neglect the $T$ and/or $\mu$ dependent 
terms in the quark and gluon propagators 
that arise from the perturbative expansion. 

Substituting Eqs. (\ref{eqn:dp}), (\ref{eqn:sigma}) and (\ref{eqn:pi}) 
into (\ref{eqn:vcjt2}) and considering the differentiation 
with respect to $\pe^2$ to be that with respect to $|\vec{p}|^2$, 
we can write down the effective potential $V(\sigma,\sigma_5;m_R)$ 
(see the Appendix of Ref. \cite{KIRIYAMA}).

In numerical calculations, since it was known that 
the temperature and chemical potential dependence 
of the quantities such as $\barq$ and $f_{\pi}$ are stable under the change of 
the infrared regularization parameter\cite{TANIGUCHI}. 
Therefore, in the first place, we fix $\ln(p_R^2/\Lambda_{\mt{QCD}}^2)=0.1$ 
and determine the value of $\Lambda_{\mt{QCD}}$ 
by the condition $f_{\pi}=93$ MeV at $T=\mu=0$ 
and in the chiral limit; i.e., $m_R=0$. 
In this case, the pion decay constant is approximately 
given by Pagels--Stoker formula\cite{PS}:
\begin{eqnarray}
f_\pi^2=4N_c\int\frac{d^4\pe}{(2\pi)^4}
\frac{\Sigma(\pe)}{(\Sigma^2(\pe)+\pe^2)^2}
\left(\Sigma(\pe)-\frac{\pe^2}{2}\frac{d\Sigma(\pe)}{d\pe^2}\right),
\end{eqnarray}
and we obtain $\Lambda_{\mt{QCD}}=738$ MeV. 
Secondly, we assume the light quarks ($u$ and $d$) are degenerate in mass and 
take the current quark mass evaluated at the renormalization point 
$\kappa=1$ GeV as $m_u(1\gev)=7{\rm MeV}$. 
Using the one-loop evolution formula, the RGI current quark mass 
$m_R$ extracted from the above-mentioned value becomes 
$m_R=7.6 \times 10^{-3}\Lambda_{\mt{QCD}}$. 
A discussion of the effective potential 
and the chiral symmetry restoration 
at high temperature and/or density can be found in Ref. \cite{KIRIYAMA}.

Figure 1 shows the phase diagram in the chiral limit 
in $n_B$-$T$ plane. 
The baryon number density $n_B$ is defined as 
$n_B=n_q/3=-(1/3)\partial V/\partial \mu$ with $V$ 
renormalized so that it has the correct free theory behavior 
at $\sigma=\sigma_5=0$. 
At $T=0$, for example, there is a mixed phase that consists of 
massive quarks with $n_B^{(-)}=1.5n_0$ and massless quarks with 
$n_B^{(+)}=4.3n_0$ where $n_0=0.17{\rm fm}^{-3}$ is normal nuclear 
matter density. The phase diagram for $m_u(1\gev)=7{\rm MeV}$ case 
is shown in Fig. 2. The critical point $E$ where 
the first order phase transition ends is found at 
$\mu_E \simeq 290$ MeV, $T_E \simeq 95$ MeV;
\begin{eqnarray}
\frac{\mu_E}{\mu_{crit}} \simeq 0.64,
~\frac{T_E}{T_{cross}}\simeq 0.70,
\end{eqnarray}
where $\mu_{crit}$ is the critical chemical potential 
at $T=0$, and $T_{cross}$ is the temperature 
where $M_\sigma$ is minimized and $M_\pi$ starts to increase at $\mu=0$. 
We have confirmed that by the finite current quark mass 
the critical point is moved from the tricritical point, which we have found 
in the previous paper\cite{KIRIYAMA}, toward larger value of $\mu$ 
and smaller value of $T$.

The values of the chiral meson masses are defined 
by multiplying the second derivative by the appropriate factor $f$:
\begin{eqnarray}
M_{\sigma}^2=f\frac{\partial^2 V}{\partial \sigma^2}\Bigg{|}_{\rm min}~,~
M_{\pi}^2=f\frac{\partial^2 V}{\partial \sigma_5^2}\Bigg{|}_{\rm min}.
\end{eqnarray}
Here, ``min'' at the end of the equations means that they are evaluated 
at the minimum of $V(\sigma,\sigma_5;m_R)$. 
In this paper, we do not examine the factor $f$; rather, 
we fix $M_\pi(T=\mu=0)$ to 140 MeV, then, $M_{\sigma}(T=\mu=0)$ turns out 
to be $668$ MeV. The $\mu$ dependence of $M_\sigma$ and
$M_\pi$ at $T=T_E$ is shown in Fig. 3. 
We find that sigma {\it almost} goes to massless at the critical point, 
while pion remains massive. 
The chiral meson masses at $\mu=\mu_E$ show behavior 
similar to Fig. 3 as functions of $T$. 
These behavior is consistent with that obtained in the linear 
the sigma model \cite{SCAVE}.

In conclusions, we investigated the phase structure and 
the chiral meson masses at the 
critical point from QCD in the improved ladder approximation. 
Using the variational approach, we calculated the effective potential 
at finite temperature and chemical potential. 
In the chiral limit, we found the mixed phase that consists of 
a low density chirally broken phase and a high density chirally 
symmetric phase when $T$ is sufficiently small. 
For explicit chiral symmetry breaking case, 
the critical point was found at $\mu_E\simeq 290$ MeV, 
$T_E\simeq 95$ MeV. The value of $\mu_E$ seems to be too large 
in light of the experiments at RHIC and LHC. 
By including strange quark mass, however, it would be reduced. 
The meson masses was obtained from the curvature 
at the minimum of the effective potential. 
We found that $M_\sigma$ almost goes to zero at the critical point 
while $M_\pi$ remains nonzero. 
The result is consistent with that obtained within 
the framework of the linear sigma model in the mean-field 
approximation\cite{SCAVE}. We have also confirmed that $\barq$ at the 
tricritical point behaves as
\begin{eqnarray}
\barq_P \sim m_R^{1/5}.\nonumber
\end{eqnarray}
Our results can be understood by the Landau-Ginzburg effective potential and 
partly support the experimental signature proposed in Ref. \cite{CSD}. 

Finally, some comments are in order. 
In this paper we did not take into account 
the screening of the gluon, that is to say, the Debye screening 
for the electric gluons. Furthermore, it has been shown that 
at finite temperature the wave function renormalization 
makes non-negligible contribution to the critical line and 
the behavior of the order parameter\cite{IKEDA}. 
They may affect the precise location of the critical point 
although the main feature of this work might not change. 
In any case, it is preferable to take into account 
the screening of the gluon, wave function renormalization 
and the effects of $s$ quark in the future work.

\section*{Acknowledgments}
The author is grateful to M. Maruyama and F. Takagi for 
valuable discussions.

\newpage
\begin{figure}
\vskip 0.2in
\epsfxsize=4in
\centerline{\epsfbox{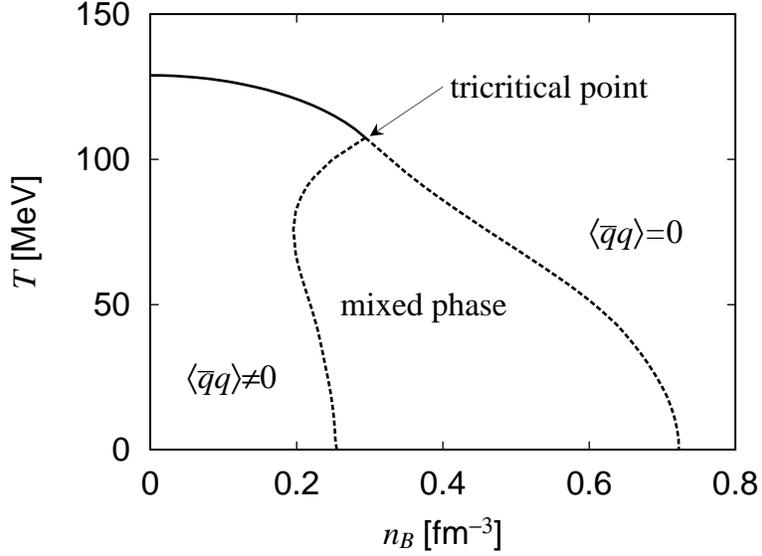}}
\vskip 0.2in
\caption{The phase diagram in the chiral limit as a function of the 
baryon density and the temperature. The solid line 
indicates the phase transition of second order 
and the dashed lines indicate that of first order.}
\end{figure}
\vskip 0.2in

\begin{figure}
\vskip 0.2in
\epsfxsize=4in
\centerline{\epsfbox{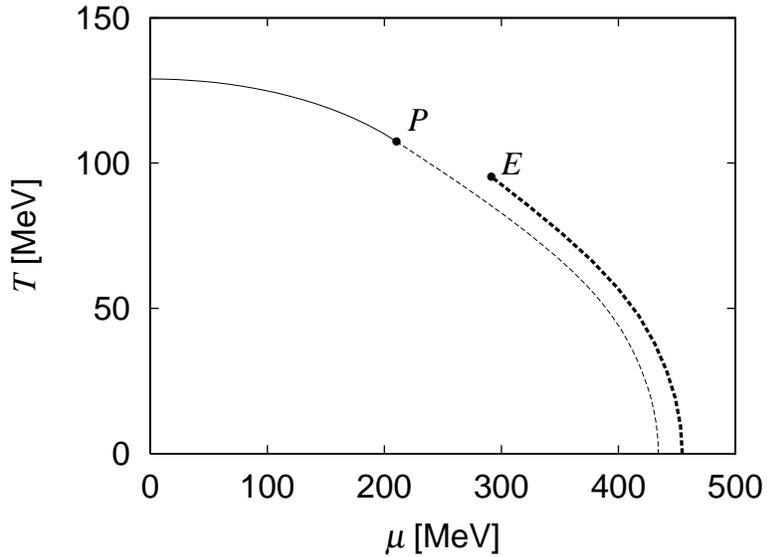}}
\vskip 0.2in
\caption{The phase diagram in $\mu$-$T$ plane. The solid line indicates 
the phase transition of second order and the dashed line indicates 
that of first order. The thin line corresponds to the chiral 
limit and the thick line corresponds to $m_R(1\gev)=7 {\rm MeV}$ case. 
The points $P$ and $E$ represent the tricritical point and 
the critical point, respectively. 
We note that the critical chemical potential 
at $T=0$ is slightly (about 2\%) larger than 
previous papers [9] because we have improved the numerical calculation.}
\end{figure}
\vskip 0.2in

\begin{figure}
\vskip 0.2in
\centerline{\epsfbox{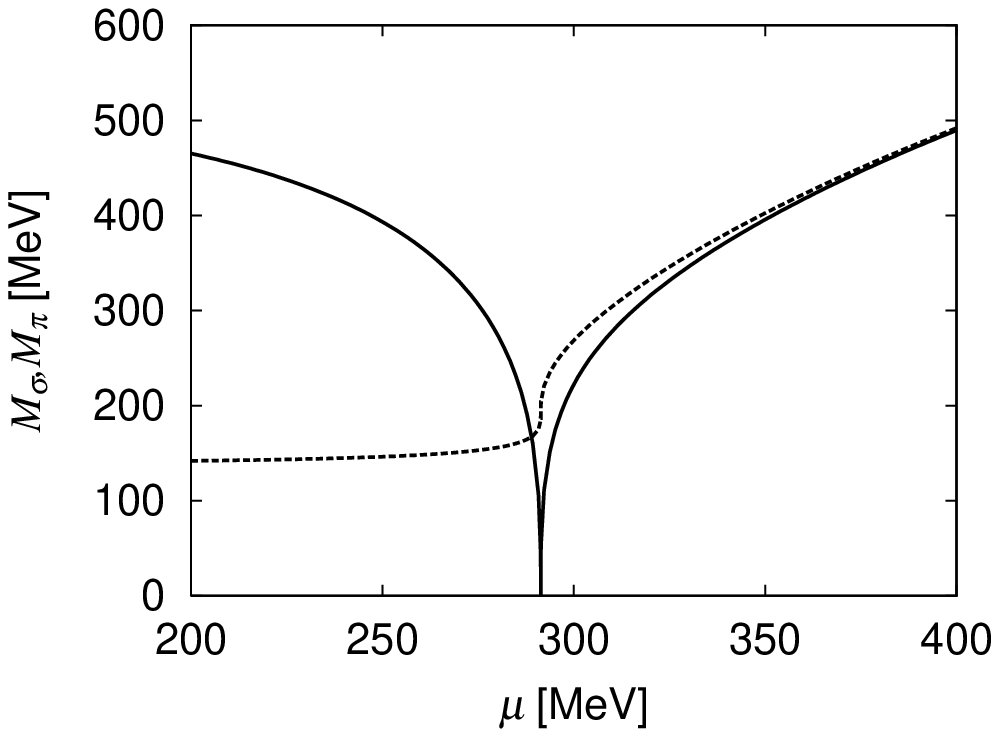}}
\vskip 0.2in
\caption{The chemical potential dependence of $M_\sigma$ (solid line) 
and $M_\pi$ (dotted line) at $T=T_E$.}
\end{figure}


\begin{references}

\bibitem{MCLERRAN}For general reviews see, for example,
J. Cleymans, R. V. Gavai, and E. Suhonen, 
Phys. Rep. {\bf 130} (1986), 217; 
L. McLerran, Rev. Mod. Phys. {\bf 58} (1986), 1021.

\bibitem{CSC}See, for example, K. Rajagopal and F. Wilczek, 
{\it The Condensed Matter Physics of QCD}, hep-ph/0011333.

\bibitem{EJIRI}S. Ejiri, Nucl. Phys. B (Proc. Suppl.) 
{\bf 94} (2001), 19. 

\bibitem{NJL}T. Hatsuda and T. Kunihiro, 
Phys. Rep. {\bf 247} (1994), 221.

\bibitem{SCAVE}O. Scavenius, \'A. M\'ocsy, I. N. Mishustin, 
and D. H. Rishke, Phys. Rev. C {\bf 64} (2001), 045202.

\bibitem{QCD}A. Barducci, R. Casalbuoni, S. De Curtis, R. Gatto, 
and G. Pettini, Phys. Rev. D {\bf 41} (1990), 1610; 
Phys. Lett. B {\bf 240} (1990), 429; 
Phys. Rev. D {\bf 46} (1992), 2203; 
A. Barducci, R. Casalbuoni, G. Pettini, and R. Gatto, 
Phys. Rev. D {\bf 49} (1994), 426; C. D. Roberts and S. Schmidt, 
Prog. Part. Nucl. Phys. {\bf 45S1} (2000), 1.

\bibitem{TANIGUCHI}Y. Taniguchi and Y. Yoshida, 
Phys. Rev. D {\bf 55} (1997), 2283.

\bibitem{HARADA}M. Harada and A. Shibata, 
Phys. Rev. D {\bf 59} (1998), 014010.

\bibitem{KIRIYAMA}O. Kiriyama, M. Maruyama, and F. Takagi,
Phys. Rev. D {\bf 62} (2000), 105008; {\it ibid.} {\bf 63} (2001), 116009.

\bibitem{IKEDA}T. Ikeda, hep-ph/0107105.

\bibitem{EP}Z. Fodor and S. D. Katz, hep-ph/0104001; 
hep-lat/0106002; hep-lat/0111064.

\bibitem{ISING}M. A. Halasz, A. D. Jackson, R. E. Shrock, 
M. A. Stephanov, and J. J. M. Verbaarschot, 
Phys. Rev. D {\bf 58} (1998), 096007; 
J. Berges and K. Rajagopal, Nucl. Phys. {\bf B538} (1999), 215.

\bibitem{CSD}M. Stephanov, K. Rajagopal, and E. Shuryak, 
Phys. Rev D {\bf 60} (1999), 114028; B. Berdnikov and K. Rajagopal, 
Phys. Rev. D {\bf 61} (2000), 105017; 
K. Rajagopal, Acta Phys. Polon. B {\bf 31} (2000), 3021.

\bibitem{BARDUCCI}A. Barducci, R. Casalbuoni, G. Pettini, and 
R. Gatto, Phys. Rev. D {\bf 63} (2001), 074002.

\bibitem{CJT}J. M. Cornwall, R. Jackiw, and E. Tomboulis,
 Phys. Rev. D {\bf 10} (1974), 2428.

\bibitem{HIG}K. Higashijima, Phys. Lett. B {\bf 124} (1983), 257;
 Phys. Rev. D {\bf 29} (1984), 1228; 
Prog. Theor. Phys. Suppl. {\bf 104} (1991), 1.

\bibitem{MIR}V. A. Miransky, Yad. Fiz. {\bf 38} (1983), 468 
[Sov. J. Nucl. Phys. {\bf 38} (1983), 280].

\bibitem{AOKI}K-I. Aoki, M. Bando, T. Kugo, M. G. Mitchard, 
and H. Nakatani, Prog. Theor. Phys. {\bf 84} (1990), 683.

\bibitem{JIK}See, for example, J. I. Kapusta, {\it Finite Temperature Field Theory} (Cambridge University Press, Cambridge, England, 1989).

\bibitem{JAIN}P. Jain and H. J. Munczek, Phys. Rev. D {\bf 44} (1991), 1873.

\bibitem{KUGO}T. Kugo and M. G. Mitchard, 
Phys. Lett. B {\bf 282} (1992), 162; {\it ibid.} {\bf 286} (1992), 355.

\bibitem{PS}H. Pagels and S. Stoker, Phys. Rev. D {\bf 20} (1979), 2947.


\end{references}
\end{document}